\documentclass[a4paper,12pt]{iopart}
\usepackage{graphicx}
\usepackage{gensymb}
\usepackage{subfigure} 
\usepackage{bm}
\usepackage{color}
\usepackage{xcolor}
\usepackage{epstopdf}
\usepackage{float}
\usepackage{soul}
\usepackage{booktabs} 
\usepackage{array} 
\usepackage{pgfplotstable}
\usepackage{pgfplots}
\makeatletter

\pgfplotsset{compat=1.16}
\begin{document}
\title[The effect of nitriding on the humidity sensing properties of a-C:H films]{The effect of nitriding on the humidity sensing properties  of  hydrogenated amorphous carbon films}

\author{{ Giuliano Frattini\dag}, Sindry Torres\dag, Leonel I Silva\ddag,  Carlos E Repetto\dag, Bernardo J G\'omez\dag and Ariel Dobry\dag}

\address{\dag Facultad de Ciencias Exactas, Ingenier\'{\i}a y Agrimensura, Universidad Nacional de Rosario -- Instituto de F\'{\i}sica Rosario, Bv. 27 de
Febrero 210 bis, S2000EZP Rosario, Argentina.}
\address{\ddag INTEMA, Facultad de Ingenier\'{\i}a, Universidad Nacional de Mar del Plata, Consejo Nacional de Investigaciones
Cient\'{\i}ficas y T\'ecnicas (CONICET), Av. Col\'on 10850, Mar del Plata, 7600, Argentina.}
\ead{bgomez@ifir-conicet.gov.ar}
\vspace{10pt}
\begin{indented}
\item[]\today 
\end{indented}

\begin{abstract}
We have studied the effect of nitriding on the humidity sensing properties of hydrogenated amorphous carbon (a-C:H) films.  
The films were prepared in two stages combining the techniques of physical deposition in vapor phase evaporation (PAPVD) and plasma pulsed nitriding. 
By deconvolution of the Raman spectrum  we identified  two peaks corresponding to the D and G  modes characteristic of a-C:H.   After the N$_2$-H$_2$ plasma treating, the peaks narrowed and shifted to the right, which we associated with the incorporation of N into the structure. 
We compared the sensitivity to the relative humidity (RH) of the films before and after the N$_2$-H$_2$  plasma treatment. The nitriding improved the humidity sensitivity measured as the low frequency resistance. 

By impedance spectroscopy we studied the frequency dependence of the AC conductivity $\sigma$ at different RH conditions. Before nitriding $\sigma(\omega)\sim A \omega^s$, it seemed to have the universal behaviour seen in other amorphous systems. The humidity changed the overall scale $A$. After nitriding, the exponent $s$ increased, and became RH dependent. We associated this behaviour to the change of the interaction mechanism between the water molecule and the substrate when the samples were nitriding.     
\end{abstract}

\vspace{2pc}
 \maketitle 

\section{Introduction}
Humidity sensors are used in many industrial applications, from 
environmental monitoring and meteorology to the determination of soil water content
in agriculture, air conditioning systems, monitoring the quality of
food, medical equipment and in many other fields \cite{farahani2014humidity}.
A  novel application of ultrafast sensors was proposed for a touch-less user interface proving to run in a whistling recognition analysis \cite{borini2013ultrafast}.  
A material to be used as a humidity sensor should change some of its properties (for example, its electrical impedance) when exposed to a humid atmosphere. The sensitivity of a sensor is the relative change of the detection property, when the RH changes in the range of interest. The requirements for such a material are: high sensitivity in order to discriminate small difference of RH, fast response and short recovery time. For a new generation of humidity sensors it would be important the compatibility with micro or nano circuits and the flexibility.

To this aim carbon films have attracted  great interest as sensitive material because they have a large sensing area  and high chemical inertness (see \cite{tulliani2019carbon} for a recent review). In particular, a recent study pointed out that hydrogenated amorphous carbon (a-C:H) films treated by N$_{2}$--H$_{2}$ plasma were sensitive to the humidity \cite{epeloa2018resistivity, frattini2021effect}. 
This films were proposed as a basis of resistivity humidity sensors.
This is a promising result due to the low cost fabrication of this material.

The goal of this paper is to analyze how plasma treatment modifies the transport properties of the films and in particular the humidity sensitive properties. For this purpose we have measured the sensitivity of the samples before nitriding, which  could not be done in Ref. \cite{epeloa2018resistivity}.

As a first step, we studied the structural changes of the films by Raman scattering. Different fitting procedures were used to determine the contribution of the G and D vibrational modes in the measured spectra.
After the plasma treatment both modes were shifted and narrowed. This was attributed to the incorporation of the N into the structure by replacing some of the C atoms. We also studied the presence of C-H bonds by IR spectroscopy. We found a very low concentration of this hydrogenated bonds.

Next, we analyzed the change of the transport properties of the films with humidity by impedance spectroscopy. The low frequency resistance changed with the humidity of the surrounding medium.  The plasma nitriding treatment improved the humidity sensing properties of the films.
 
In order to understand how the humidity changed the electrical transport of the films, we studied the Nyquist plots of Impedance Spectrum and the frequency dependence of AC conductivity at different humidity. The Nyquist plots had a semicircular-type behaviour.   A straight line-type impedance characteristic of the diffusion of ions
across the interface was not observed. Therefore, we associated the RH dependence to the bulk properties of the films.

As regards the dependence of the frequency of the real part of the conductivity, its behaviour changed after  the  nitriding process. Before  it,  they  had a  kind  of  universal behaviour which resembled the one  found  in  some  amorphous  semiconductors for the AC conductivity at different temperatures \cite{baranovski2006charge}. When  the films were nitrated, the slope of the $\sigma(\omega)$ decreased with  the  humidity  and  all these curves converged at high frequency.  We understand that the improvement of the sensing properties with the humidity was related with a change in the interaction mechanism between the water molecules and the film.

\section{Synthesis and characterization of a-C:H films}

The carbon films were deposited by the electron-beam physical vapour deposition technique, in which small graphite bars were bombarded with an electron beam that was generated in a tungsten filament powered by an external source at $150$ mA. This procedure was carried out in a high vacuum chamber at $10^{-6}$ Torr. The evaporated gas was deposited  over printed circuit boards made of synthetic resin FR2, predesigned to measure electric transport.

Later, the sample was placed in a plasma nitriding reactor originally designed to treat the surface of steels. The purpose of this procedure was to dilute the sample to increase the resistance and to change the microstructure of the film by interaction with the plasma.
The reactor consisted of an 8-litre AISI 304 stainless steel vacuum chamber connected to ground potential. Inside the chamber, there was an electrode, also made of AISI 304 stainless steel, connected to a power supply that was able to generate a square-wave signal of up to 700 V at a frequency between 0 and 1 kHz. The duty cycle can also be changed.
A pre-vacuum of 0.001 Torr was created in the chamber, which was then back filled to 1 Torr with a mixture of 50\% nitrogen and 50\% hydrogen. We conducted our experiments at a constant temperature of 250\degree C, with an on/off ratio of 50\%/50\% at a frequency of 100 Hz. The applied voltage was 550 V. Pressure, current, voltage and flux were kept constant during the ion nitriding treatments. The treatments lasted 5 min; measured from the moment the sample temperature reached 50\degree C.

After the nitriding process we named the samples as a-CN:H to distinguish them from the ones we had before the plasma treatment which we had called a-C:H. By spectroscopic  ellipsometry measurements we determined that the thickness of the sample had been strongly reduce after the plasma treatment. It was $70$ nm for a-C:H and became $20$ nm for a-CN:H. 

The Raman spectrum shape fitting is a widely used method to study the detail bonding structure of carbon films.
Therefore, we obtained the Raman spectra in order to characterize the structure of the deposited films.

The spectra were acquired in a Renishaw In Via reflex system equipped with charge-coupled device (CCD) detector of 1040$\times$256 pixels. A 514 nm diode laser (50 mW) was used as excitation source in combination with a grating of 2400 grooves/mm and slit openings of 65 $\mu$m, which yield a spectral resolution of about 4 cm$^{-1}$. The laser power was kept below 10\% to avoid sample damage. A 50$\times$(0.5 NA) with shorter working distance (210 $\mu$m) Leica metallurgical objective was used in the excitation and collection paths. Spectra were typically acquired in 10 s with at least 10 accumulations.

A region of the characteristic spectrum of a diamond-like carbon (DLC) structure was observed. Once the background was subtracted, two main bands were observed, corresponding to the D and G modes. They were located at about 1400 and 1550 cm$^{-1}$ wavenumbers, respectively. 

The sample had an area of 1 cm$^2$ and could be inhomogeneous. Therefore, Raman spectra were taken at different points.  The spectra were fitted by taking into account different possibilities \cite {Tai2009,Yuan2017}. Among the different alternatives, we chose to use  Gaussian functions  for both peaks or a Lorentzian and a Gaussian functions for each peak. We also used a BWF function for D peak and a Lorentzian function for G peak.
 
\begin{figure}[!ht]
\begin{center}
\includegraphics[trim = 0cm 0cm 0cm 0.0cm, clip, width=18pc]{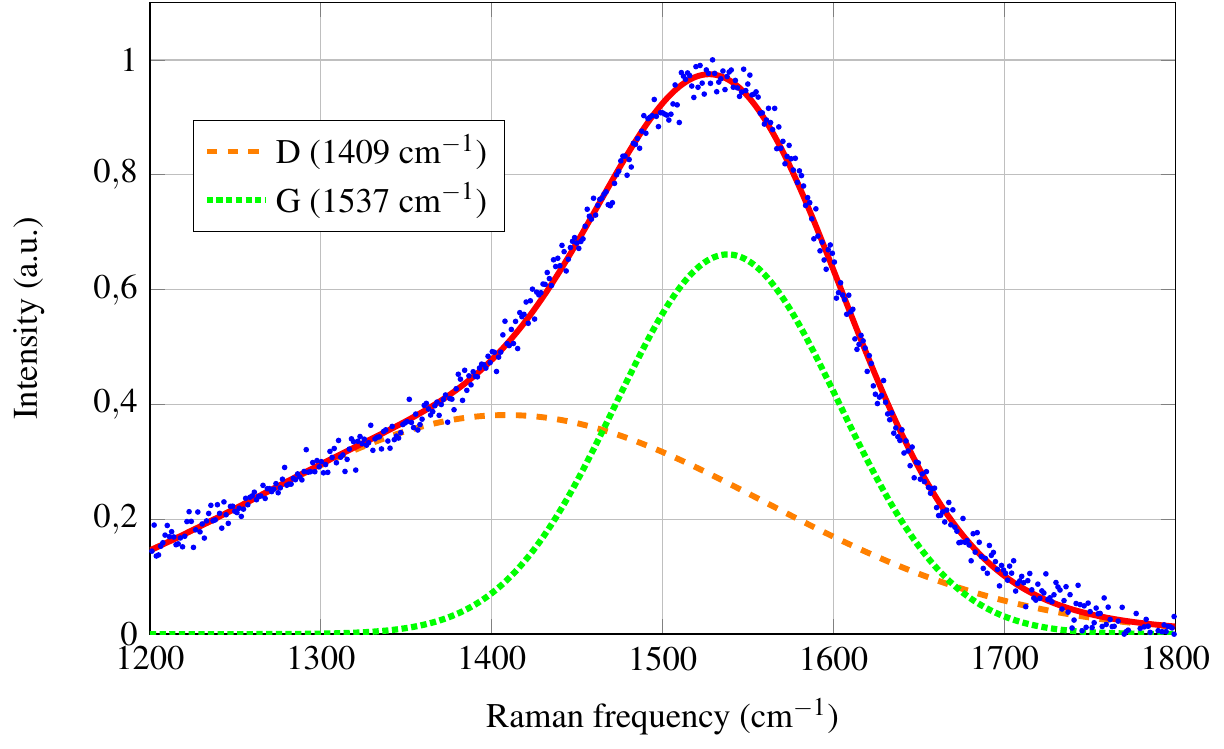}
\includegraphics[trim = 0cm 0cm 0cm 0.0cm, clip, width=18pc]{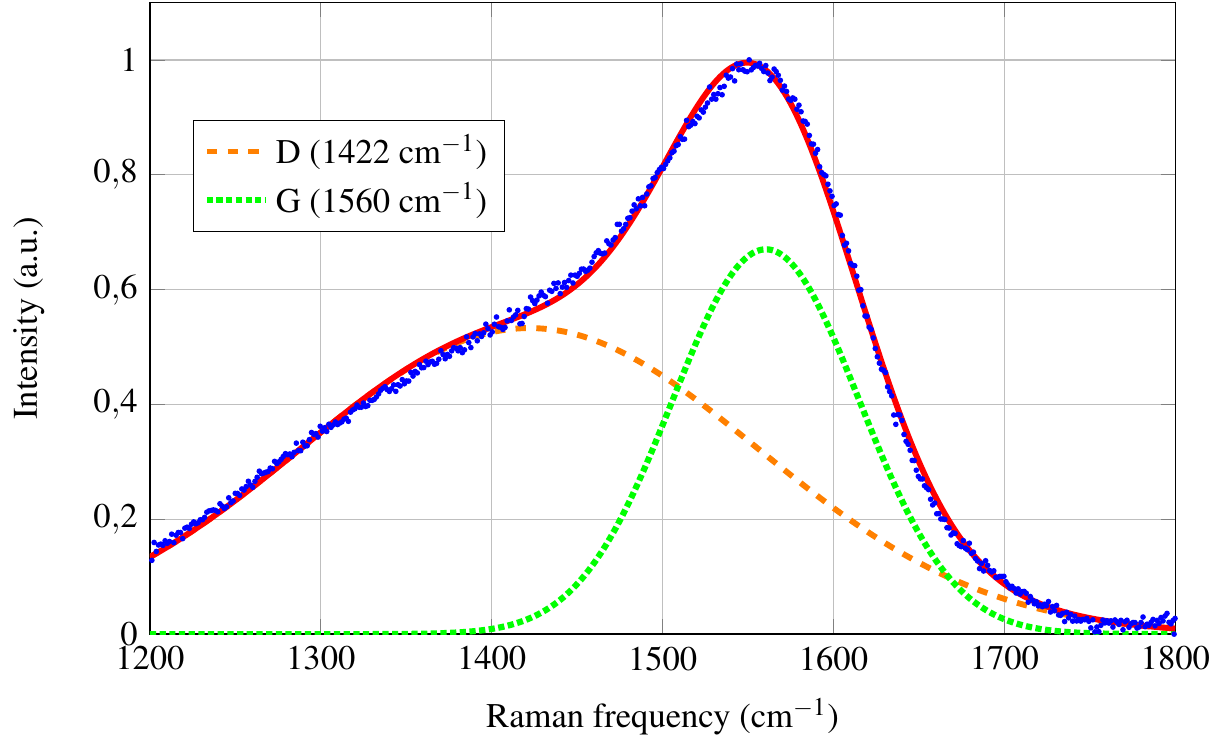}
\caption{In blue points, the Raman spectra for a-C:H (left) and a-CN:H (right) samples. Both graphics show a deconvolution on two Gaussian peaks corresponding to D and G modes. In solid red line, the sum of the two gaussians.}
\label{fig:before}
\end{center}
\end{figure}

As an example, we can see in Fig. \ref{fig:before} a deconvolution  with Gaussians functions for D and G peaks for a-C:H and a-CN:H samples.

The values of the parameters obtained from the different  fittings are shown in Table \ref{table1}. They are average values from the spectra obtained at different points of the sample. 
  
\begin{table}[ht!]
\centering
\begin{tabular}{|c|c|c|}
\hline
Parameter & a-C:H & a-CN:H  \\ 
\hline
$x_D$ & (1409$\pm$12) cm$^{-1}$ & (1425$\pm$5) cm$^{-1}$ \\ \hline
$s_D$ & (153$\pm$6) cm$^{-1}$ & (143$\pm$4) cm$^{-1}$  \\ \hline
$x_G$ & (1537$\pm$1) cm$^{-1}$ & (1554$\pm$1) cm$^{-1}$  \\ \hline
$s_G$ & (73$\pm$3) cm$^{-1}$ & (63$\pm$2) cm$^{-1}$  \\ \hline
$I_D/I_G$ & (0.5$\pm$0.1) & (0.7$\pm$0.2) \\ \hline
$A_D/A_G$ & (1.1$\pm$0.3) & (1.7$\pm$0.6)  \\ \hline
\end{tabular}
  \caption{ Results for the Raman D and G peaks adjustment parameters for a-C:H and a-CN:H samples.}
  \label{table1}
\end{table}
Here, $x_D$ and $s_D$ are the position and FWHM of the D peak, $x_G$ and $s_G$ are those corresponding to the G peak. $I_D/I_G$ and $A_D/A_G$ are the ratios of intensities and areas between D and G peaks, respectively.

From the values in Table \ref{table1}, we can conclude that the plasma treatment of the sample caused peaks D and G to shift to the right by 1.1\%, the peaks narrowed by 6.4\% and 13.9\%, and the ratio of intensities and areas increased by 35.4\% and 53.3\%, respectively.

It is well known that the plasma treatment produces the incorporation of nitrogen in the treated samples. Previous studies on a-C$_{1-x}$N$_x$:H  have determined the shift of $x_D$ and $x_G$  and the increasing of  $A_D/A_G$  as given by a Gaussian fit, as a function of the $x$ content of N  \cite{godet2005optical, franceschini2002growth}. By comparing our results for the shift of the G peak and the relation of the areas, we estimated that there could be an incorporation of nitrogen in the sample structure of the order of $10\%$.

The increase of $I_D/I_G$ and the shift of $x_G$ were related to the increase of the aromatic $sp^2$ zones \cite{ferrari2000interpretation}. This structural change produced an increase of the DC conductivity, which was mainly associated with an increase of the density of states at the Fermi level $N(E_F)$ \cite{godet2005optical,schwan1994microstructures}. 

\begin{figure}[!ht]
\center
\vspace{0.6cm}
\includegraphics[trim = 0cm 0cm 0cm 0.0cm, clip, width=18pc]{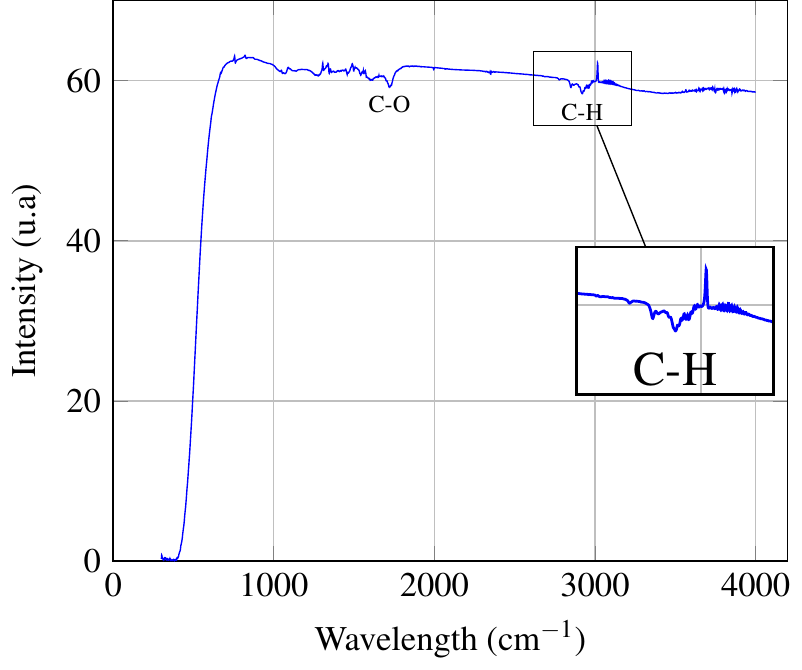}
\caption{The IR spectrum of the film before plasma treatment at room temperature.}
\label{fig:ir}
\end{figure}

Independent information on the structure of the film was obtained by IR spectroscopy. We used the Spectrometer One by Perkin-Elmer to obtain the result shown in Fig. \ref{fig:ir}, which corresponds to the sample before the plasma treatment.

The peaks were attributed to different configurations of sp orbital. For the peak at 2920  cm$^{-1}$ the stretch vibration C-H was assigned $sp^3$\cite{veres2002ir}. For the peak 1720 cm$^{-1}$ it was attributed to vibration C-O, indicating that the film had hydrogen in a very low concentration.   

\begin{figure}[!ht]
\center
\vspace{0.6cm}
\includegraphics[trim = 0cm 0cm 0cm 0.0cm, clip, width=25pc]{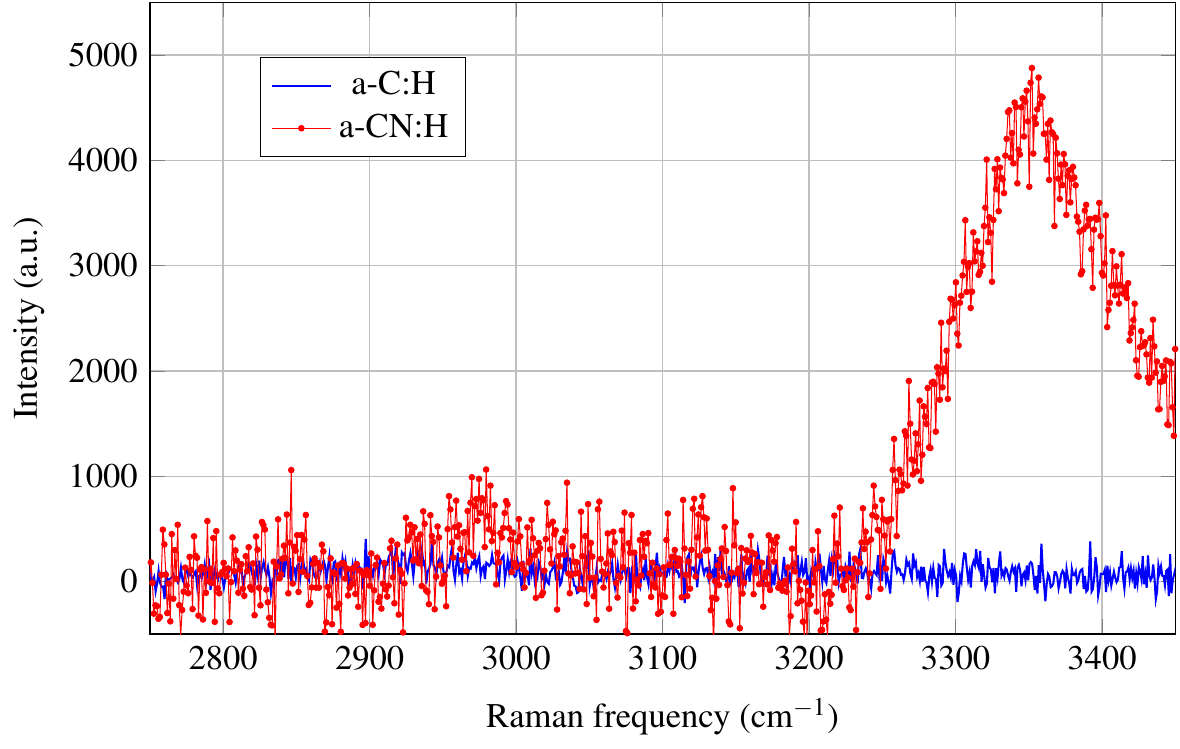}
\caption{The Raman spectrum of the film before (a-C:H) and after (a-CN:H) plasma treatment at room temperature.}
\label{fig:ramanr}
\end{figure}

N incorporation into the a-C:H structure produces important changes in the IR and Raman spectra, as has been shown in previous studies \cite{franceschini2002growth,kaufman1989symmetry}. In Fig.  \ref{fig:ramanr} we show the Raman spectrum of the sample before and after the plasma treatment.
The existence of a peak in the 3360  cm$^{-1}$  which is assigned to a NH stretching mode can be clearly seen in the signal corresponding to the a-CN:H  sample. The appearance of this mode after N-doping has also been seen by IR spectroscopy \cite{kaufman1989symmetry,Kipkemboi2003}.
This result would suggest that most of the  nitrogen atoms are sitting at the terminal sites of the amorphous network and therefore hydrogenated \cite{franceschini2002growth}.

In summary, the effect of nitriding was a reduction of the thickness of the sample and the incorporation of N into the structure.
The resulting microstructure is similar to the one seen in the previous work where the samples were prepared by decomposition of (C$_2$H$_2$, N$_2$)  gas mixtures in a distributed electron cyclotron resonance (DECR) plasma \cite{godet2005optical}. It is a transition from a diamond-like a-C:H films towards a graphite-like N-rich a-C$_{1-x}$N$_{x}$:H alloy. Note that the precise microstructure of  a-C$_{1-x}$N$_{x}$:H is still under discussion.

In the next section we will show that the way conductivity changes with RH also changes when nitriding.

\section{Humidity dependence of the AC transport properties}

In this section we are going to study the evolution of the complex impedance of the films as the RH varies. For this purpose, we used a controlled humidity and temperature chamber. Inside we placed a commercial reference sensor SHT 71 (www.sensirion.com) to obtain the inner temperature and humidity. 

Let us start by describing the measurement equipment.  A valve was attached to the chamber for dry air intake. In addition, a heating resistor was placed inside. In this way, the humidity in the chamber decreased by introducing dry air or increased by evaporation of water droplets placed over the heating element. Thus, a humidity variation between 15 \% and 100 \% is achieved and with a temperature variation of less than 2\degree C.

A Lock-in SR530 amplifier was used to perform the electrical characterization of the amorphous carbon film. From the measurement of the current flowing through the film, the impedance was obtained by keeping the frequency fixed and varying the humidity or vice versa.

The acquisition of the measurements data was carried out through a PC that was connected to the Lock-in amplifier by means of an SR280 interface. Using a  Python program, complex current measurements taken by the Lock-in amplifier were processed and transformed into complex impedance values. Therefore, the frequency, humidity and temperature values were registered and controlled inside the chamber at the same time.

First, the humidity inside the chamber was set to a value and then measurements were made by sweeping the frequency values from 1000 Hz to 100 kHz with a certain step. 

In a previous work \cite{epeloa2018resistivity}, we showed that after nitriding the low frequency impedance changed with RH. We also showed that the RH dependence of impedance was dominated by its resistive component. Then, the films could be the basis of resistivity humidity sensors.

\begin{figure}[!ht]
\center
\includegraphics[width=18pc]{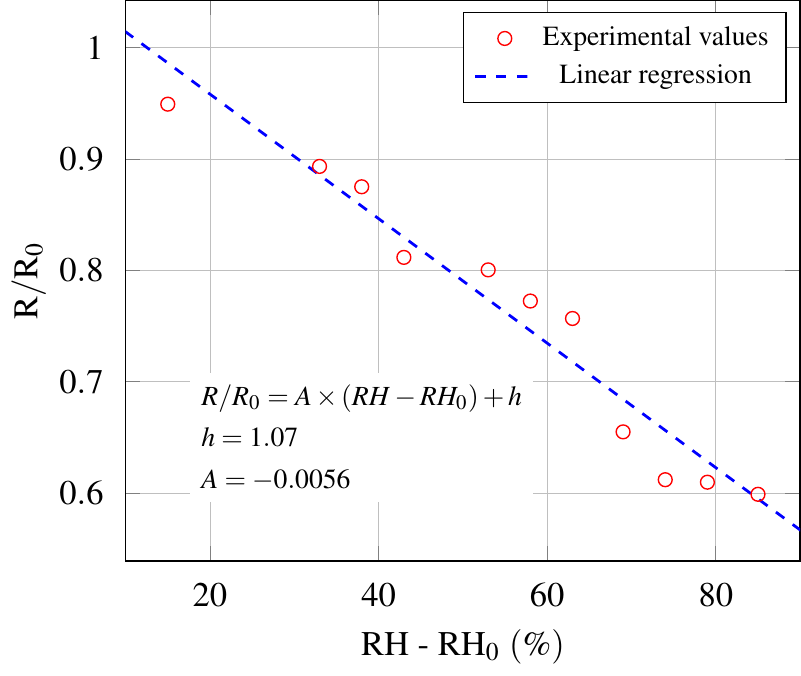}
\includegraphics[width=18pc]{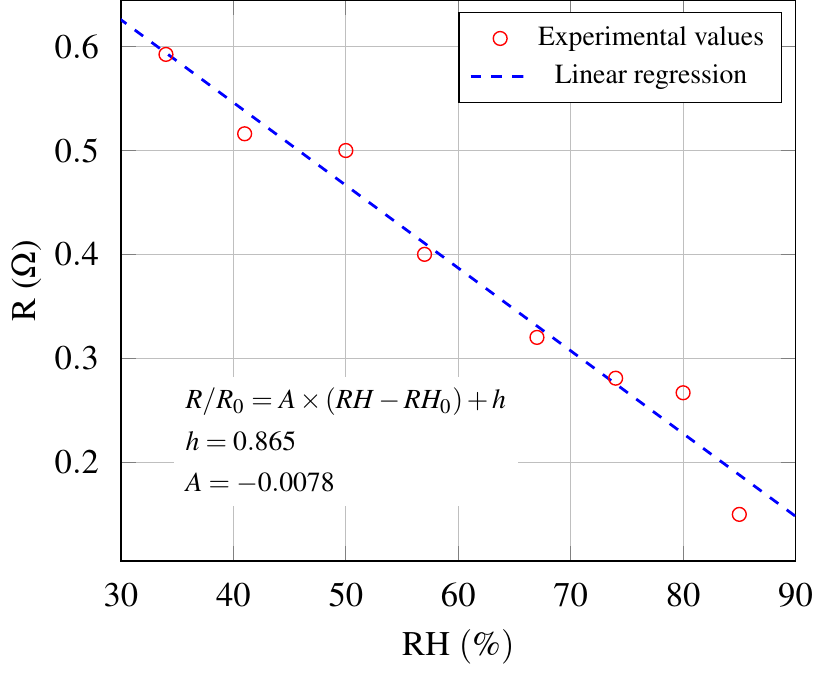}
\caption{Resistance of a-C:H (left) and a-CN:H (right) films.}
\label{fig:realpost}
\end{figure}

Therefore, we started by studying the low frequency resistance (essentially its DC value) as the sensible property. $R$ was obtained as the real part of the impedance $Z$. In Fig.
\ref{fig:realpost} we show how $R/R_0$ changed with the relative humidity ($RH$) of the surrounding media, where $R_0$ was a reference resistance at $RH_0 = 15\%$.

From Fig. \ref{fig:realpost} it can be seen that the slope of the linear regression of the a-CN:H film became greater than the one of a-C:H. In other words, the sensitivity of the sample increased after the plasma treatment.

Let us undertake a comparison of the sensitivity of a-C:H and a-CN:H samples. By defining the sensitivity as $S = \frac{R_{15\% RH}-R_{100\% RH}}{R_{15\% RH}}$, we measured it and determined that it changed from $0.40$  to $0.85$ when N was incorporated. We arrived at the conclusion that:
\begin{itemize}

\item a-C:H samples were already humidity sensitive, and 
\item The incorporation of N improved the sensitivity of the films.

\end{itemize}
Therefore both a-C:H and a-CN:H films could be used as a basis of an humidity sensor but the last one has better properties.

We have not measured this response time because it requires a different experimental device. We will do this determination in a future work. However, we know that the response time of the film is at most that of the commercial sensor SHT 71 (www.sensirion.com) used to determine the RH. This statement is supported by observing that in the acquisition of impedance spectroscopy data, no steps (plateaus) are recorded. That is, every time the commercial sensor changes the humidity record, so does the impedance of the sample.

In the following we are studying how the humidity sensing properties change by nitriding. We have used a power full technique of impedance spectroscopy. In a first step we studied the Nyquist plot  corresponding to $-Z''$ vs $Z'$. Then, we analyzed the frequency dependence of the complex conductivity ($\sigma$). 

\subsection{Humidity dependence of the Impedance}
\label{Zhum}
\begin{figure}[!ht]
\center
\includegraphics[width=36pc]{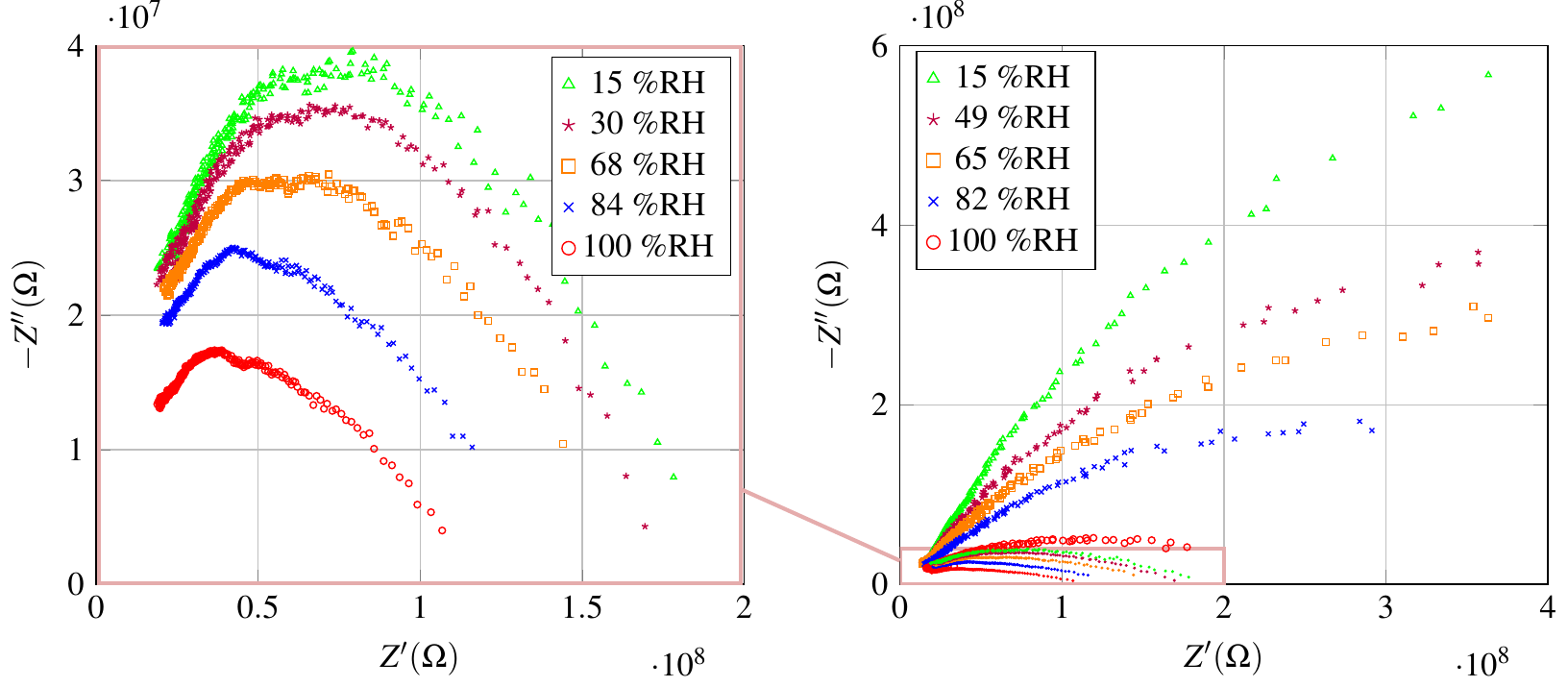}

\caption{\label{fig:Nyquist} Imaginary part vs. real part of the complex impedance due to different relative humidity conditions, for  a-C:H and a-CN:H films (right). On the left we zoomed in on the pink rectangle on the right in order to distinguish the behaviour of a-C:H.}
\end{figure}

Following the same measurement procedure described above, we swept the frequency in the previously stated range and then changed the humidity in the chamber. In Fig. \ref{fig:Nyquist} we plotted the imaginary part of the complex impedance ($Z''$) as a function of the real part ($Z'$), before  and after the plasma treatment: these are the so called Nyquist plot of impedance \cite{macdonald2005impedance}. At the right figure the results of both samples are superposed whereas at the left we zoomed in the a-C:H samples in order to show their behaviour. In these curves the frequency is taken as a parameter and increases counterclockwise.

It is clearly seen that while the real part of the impedance increases by a factor of two, after the plasma treatment, the imaginary part increases by an order of magnitude, approximately. From the behavior of the impedance shown in Fig. \ref{fig:Nyquist}, it can be inferred that the film behaves as a succession of resistors and capacitors in parallel.
It is important to note that the graphs have different scales in the imaginary axis in order to compare its behaviour.
On the other hand, we want to emphasize that we have not observed a linear behavior at low frequencies, characteristic of the response to the presence of the called Warburg resistor. I.e. a constant phase element associated with a charge-transfer resistance. This would be a signal of an additional conduction due to ionic layer formed on the surface of the sample \cite{yao:2012}.  \\
 In order to known the transmission mechanism of the current, we will study in the next section the behavior of conductivity as a function of frequency.

\subsection{Frequency and humidity dependence of the conductivity}
As we stated before, we will now study the conductivity as a function of frequency, for different RH values. From the measured values of $Z'$ and $Z''$, the values for the real ($\sigma'$) and imaginary ($\sigma ''$) parts of the admittance can be obtained:
\begin{equation}
\sigma' = \frac{e}{A} \frac{Z'}{Z'^2 + Z''^2} \qquad , \qquad \sigma'' = \frac{e}{A} \frac{Z''}{Z'^2 + Z''^2} \,\;,
\label{eq.sigma}
\end{equation}
where $e$ and $A$ are the thickness and the area of the sample, respectively. We do not include $A$  in the following figures because it is the same for all the samples and result in a constant shift of the y-axis.

\begin{figure}[!ht]
\center
\includegraphics[trim =0.cm 0.cm 0.cm 0.0cm, clip, width=18pc]{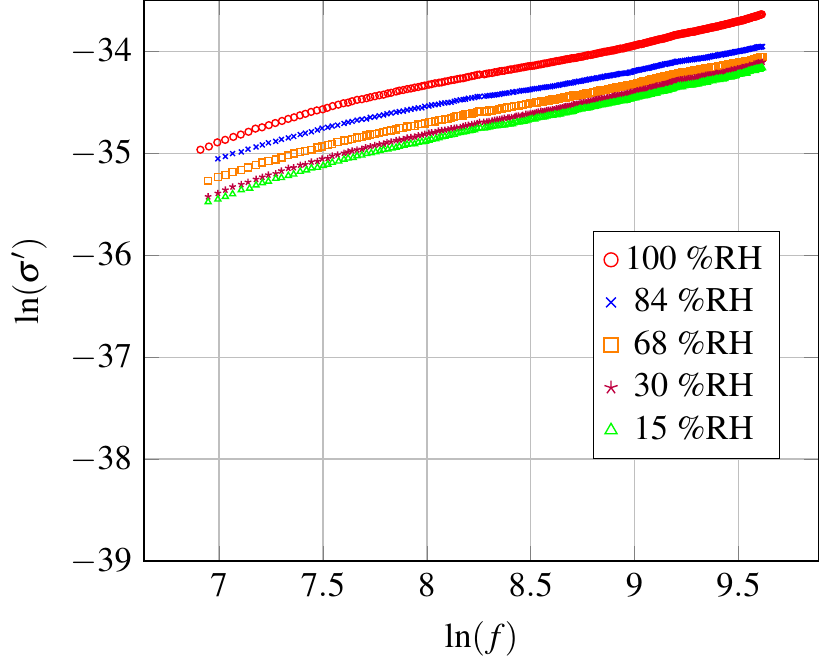}
\includegraphics[trim =0.cm 0.cm 0.cm 0.0cm, clip, width=18pc]{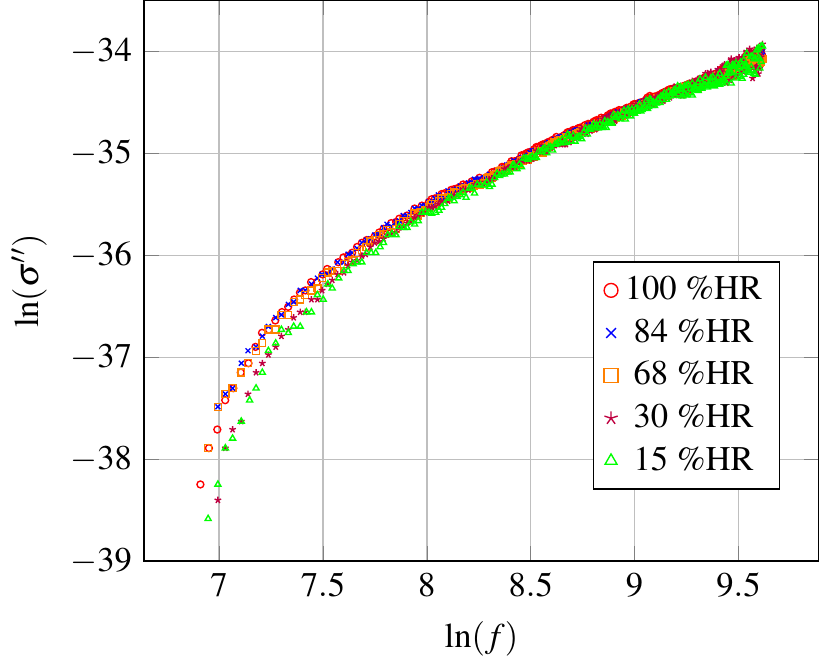}
\caption{\label{fig:lncondpre}Log-log plots of the real (left) and imaginary (right) part of the complex conductivity vs. frequency for an a-C:H film at different relative humidity conditions.} 
\end{figure}

\begin{figure}[!ht]
\center
\includegraphics[trim =0.cm 0.cm 0.cm 0.0cm, clip, width=18pc]{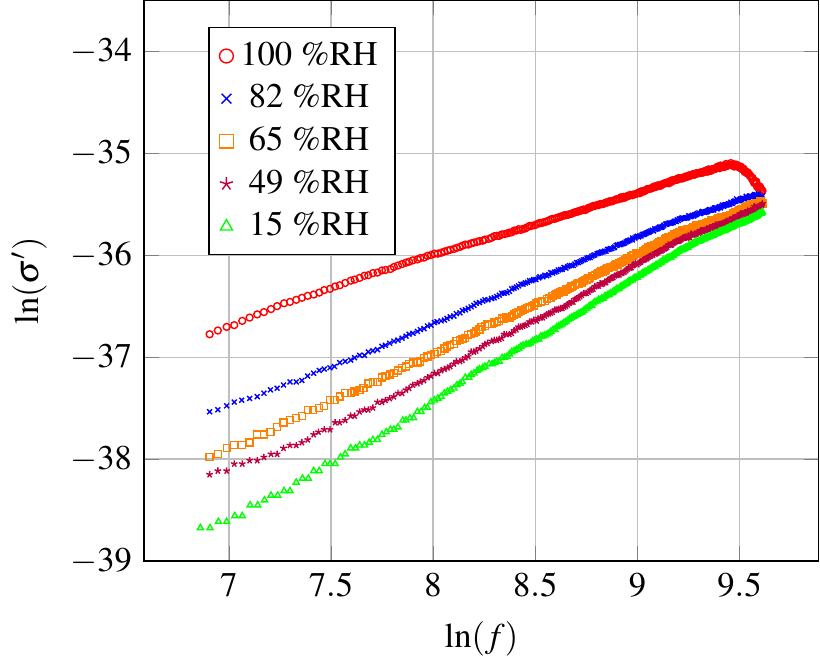}
\includegraphics[trim =0.cm 0.cm 0.cm 0.0cm, clip, width=18pc]{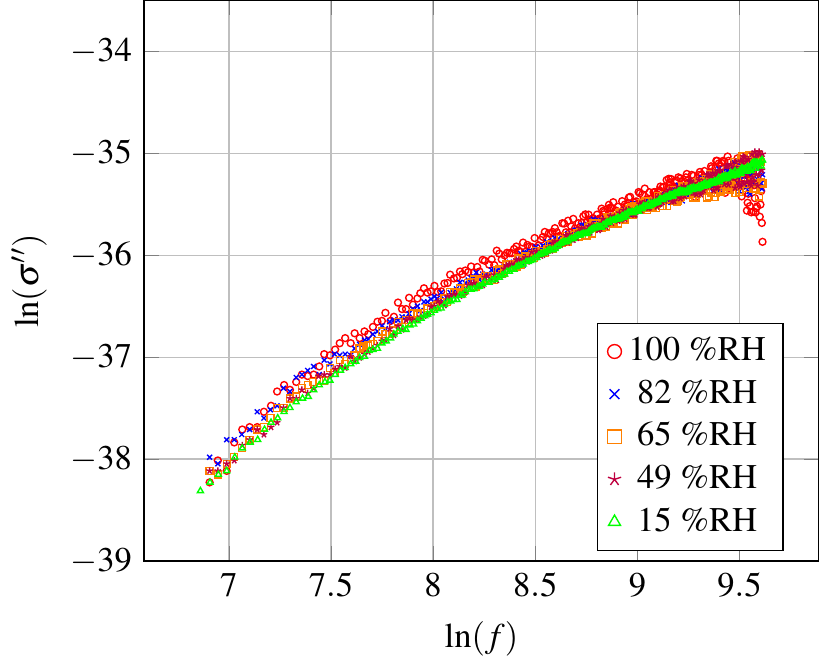}
\caption{\label{fig:lncondpost}Log-log plots of the real (left) and imaginary (right) part of the complex conductivity vs. frequency for an a-CN:H film at different relative humidity conditions.}
\end{figure}
 In  figures  \ref{fig:lncondpre}  and  \ref{fig:lncondpost}  we  can  see  the  real  (left) and  imaginary (right)  parts  of  the  conductivity 
 before and after the plasma treatment, respectively.
In both figures, the real part shows greater dependence on the humidity conditions than the imaginary part. 

For the imaginary part, the plasma treatment causes the curves at different RH to separate slightly and vary in a little smaller range as a function of the frequency. On the contrary, the real part shows a greater variation on the humidity conditions. For the a-C:H films, all the curves show similar slopes, they are almost parallel. It can be seen that they have a type of universal behavior that resembles the ones found for AC conductivity of some amorphous semiconductors at different temperatures \cite{baranovski2006charge}. This changes noticeably for the a-CN:H films, showing different slopes: the higher the humidity, the higher the conductivity but the lower the slope as a function of the frequency. Furthermore, it is important to note that all these curves converge at the same point at high frequencies. We note that this behavior resembles the one shown by thermal conductivity as a function of temperature, see for example Ref. \cite{Cahill92}. The authors of this work show that the vibrations of the lattice of these disordered crystals are essentially the same as those of an amorphous solid.

\section{Conclusion}
In summary, nitriding a-C:H samples result in a-CN:H where N is incorporated into the structure. The electronic transport properties were affected by this treatment. In particular, the resistance of a-CN:H was more sensitive to the change of the RH of the surrounding medium that the one of a-C:H.
In summary, nitriding a-C:H samples result in a-CN:H where N is incorporated into the structure. The electronic transport properties were affected by this treatment. In particular, the resistance of a-CN:H was more sensitive to the change of the RH of the surrounding medium than the one of a-C:H. As we stated in our previous work [4], the sample remains humidity sensitive for at least nine month. It is worth mentioning that the sample, during this period, was subjected to a large number of cycles of changes in the condition of relative humidity.

By AC spectroscopy we have observed that the evolution of $\sigma$  with the frequency for different RH was different before and after nitriding. We saw that water molecules adsorbed on the surface had a different effect into the conduction mechanism of both samples. We propose that the humidity affects the bulk conduction of the samples. Conduction by an ionic layer does not seem to be the main mechanism of current transport.    
\section{Acknowledgment}
We acknowledge V. Roldan, G. Baranello, H. Rindizbacher and Javier Cruce\~no for their essential contributions to resolving all the technical issues to undertake the present work. This work was supported by CONICET and UNR. Leonel Silva acknowledges Agencia Nacional de Promoción Científica y Tecnológica, Fondo para la Investigación Científica y Tecnológica (PICT16‐3633) for financial support.

\section*{References}
\bibliographystyle{iopart-num}
\bibliography{biblio}
\end{document}